# Spin dynamics in a compound semiconductor spintronic structure with a Schottky barrier

Semion Saikin[1-3], Min Shen[1], and Ming-Cheng Cheng[1]

[1]Center for Quantum Device Technology
[1]Department of Electrical and Computer Engineering
Clarkson University, Potsdam, NY 13699

[2]Department of Physics, University of California San Diego,
La Jolla, California 92093-0360

[3] Department of Physics, Kazan State University, Kazan, Russian Federation 420008

*Abstract*

We demonstrate theoretically that spin dynamics of electrons injected into a GaAs semiconductor structure through a Schottky barrier possesses strong non-equilibrium features. Electrons injected are redistributed quickly among several valleys. Spin relaxation driven by the spin-orbital coupling in the semiconductor is very rapid. At $T = 4.2$ K, injected spin polarization decays on a distance of the order of $50 - 100$ nm from the interface. This spin penetration depth reduces approximately by half at room temperature. The spin scattering length is different for different valleys.

**Introduction.**

Electrical spin injection into a non-magnetic semiconductor structure is one of the most complicated issues in design of semiconductor spintronic devices [1-3]. High efficiency of spin injection in magnetic/nonmagnetic semiconductor structures has been demonstrated in the diffusive transport regime [4]. Also, in the ballistic transport regime, spin filtering with a magnetic semiconductor can lead to nearly 100% spin injection [5]. However, at the present stage spin-dependent properties of most of magnetic semiconductors are limited by the low temperature regime only, that strongly restricts their device application. Ferromagnetic metal contacts possess much higher Curie temperature and are more attractive for the application in room temperature spintronic devices. But, in conventional ohmic metal/semiconductor contacts large conductance mismatch [6] prevents efficient injection of spin polarized carriers. One of the solutions for the problem is to utilize a tunneling barrier to increase interface resistance between a ferromagnetic metal source and a semiconductor structure [7,8]. This has been demonstrated experimentally in several different designs [9-11]. Among these designs, the Schottky contact attracts more attention because of several reasons. Firstly, the barrier naturally appears at the metal/semiconductor interface [12]. Secondly, it can be easily modified by an interface doping layer to increase tunneling current. Additionally, spin polarized electrons can be injected into a semiconductor up to the room temperature [10,13], and a device structure can be scaled down to the size of modern electronic devices.

Several different theoretical models have been utilized lately to describe spin injection through a Schottky barrier [7,8,14,15]. However, most of these approaches are based on the assumption of quasi-equilibrium transport. Using an ensemble Monte Carlo simulation method [15], we have demonstrated that near the Schottky barrier charge and spin dynamics of electrons are far from equilibrium. Strong electric field at the interface leads to redistribution of electrons among several valleys of conduction band in the vicinity of the barrier. Due to the coupling of electron spin with its spatial motion (spin-orbital coupling) [16], spin polarization in semiconductor dissipates very rapidly on a deep submicron length scale that is several times shorter than the energy relaxation length. The spin scattering lengths are different in the valleys of different symmetry.

**Model**

The device modeled is a spin Schottky diode of 1 μm length that consists of a ferromagnetic metal contact and a semiconductor channel. The semiconductor channel is modulated-doped. The doping profile together with the simulated potential profile for a reverse bias $V_{bias}$ = 2 V is shown in Fig. 1. The 30 nm $n^+$-layer of the channel at the interface reduces the width of the Schottky barrier and increases the efficiency of electron tunneling from the ferromagnetic source. On the right side of the device, a highly doped layer of 250 nm width is used to produce an ohmic contact at the drain electrode. The remaining part of the channel is homogeneously n-doped. The doping concentration is equal to $2.5 \cdot 10^{18}$ cm$^{-3}$ and $2.5 \cdot 10^{16}$ cm$^{-3}$ in the $n^+$ and n regions, respectively. The magnetization of the ferromagnetic contact is assumed orthogonal to the metal/semiconductor interface. Such a structure can be considered as a simplification of an existing spintronic device, spin-light-emitting diode (spin-LED) [17]. We emphasize that the simplified structure is chosen to address spin dynamics near the Schottky barrier rather than functionality of a spin-LED. The latter is more complex and should include electron capture in the quantum well and exciton recombination processes. We use Fe as the source contact of spin polarized electrons and GaAs as the device channel. These materials are widely utilized in spintronic studies [1]. The Schottky barrier height in this case is taken as $V_B$ = 0.72 eV [18].

To simulate transport properties, we utilize the ensemble Monte Carlo simulation approach [19]. It is a powerful tool that can account for details of a device structure and address transport properties in different regimes. It has been successfully applied to study hot electron transport in bulk GaAs in [20]. The approach has also been used to study the tunneling phenomenon in ballistic electron emission microscopy [21] and also in Schottky contacts [22,15,23].

In the model, we consider electron transport in the three non-equivalent valleys ($\Gamma$, L and X), where electron interactions with impurities, acoustic phonons and optic polar phonons are accounted for intra-valley scattering and interaction with optic non-polar phonons for inter-valley scattering [24]. Other conventional details of the approach can be found in [19]. Here we emphasize on some assumptions used for simulation of tunneling through the barrier and electron spin dynamics in the valleys.

In general, to describe spin injection from a ferromagnetic metal into a non-magnetic semiconductor, one has to start with first-principle models [25]. The difficulty is that the metal conduction band structure is very complex [26], and the effective mass approximation [27], used for semiconductors, should be applied with much caution to avoid non-realistic results. Within single particle approximation, the tunneling current density from a metal contact into semiconductor is [28]

$$j = \frac{qg_\sigma}{(2\pi)^3} \int f_m(E)(1-f_s(E))v_x T(E) d^3k, \qquad (1)$$

where $q$ is an electron charge, $g_\sigma$ accounts for spin degeneracy, $E$ is the electron energy, $v_x$ is the x-component of the electron velocity, $f_m(E)$ and $f_s(E)$ are distribution functions in the metal and semiconductor respectively, and $T(E)$ is the transmission coefficient. Integration is taken over the components of the electron wave vector, **k**. The transmission coefficient, $T(E)$, in Eq. (1) accounts for matching of materials at the metal/semiconductor interface and also for tunneling through the potential barrier [12] inside the semiconductor. There is no good analytical model for electron transmission through the ferromagnetic metal/semiconductor interface. Therefore, we try to separate its effect from the barrier tunneling. Firstly, we assume that only the majority electrons can be injected through the interface into the semiconductor. Electrons from the $\Delta_1$ conduction band of the ferromagnetic iron (for the band structure see [26]) can penetrate into a semiconductor for a much longer distance than other bands [25] because of the better matching on the interface. The minority conduction band of $\Delta_1$ symmetry is far above the barrier, and its contribution in spin injection is negligible. The next assumption is that the transmission coefficient through the material interface can be factorized from the barrier tunneling coefficient which is described by the WKB approximation [29]. The total tunneling coefficient can therefore be written as

$$T(E) = T_{\text{int}1} e^{-\frac{2\sqrt{2m^*}}{\hbar} \int_0^X \sqrt{V_B(x) - E_x} \, dx} T_{\text{int}2}, \qquad (2)$$

where $m^*$ is the electron effective mass in the semiconductor, and $V_B(x)$ is the barrier profile. $T_{\text{int}1}$ and $T_{\text{int}2}$ account for metal/semiconductor and semiconductor/semiconductor interfaces of the barrier, respectively. It can be shown easily that Eq. (2) is exact for a thick rectangular barrier [25]. One can try to account for the interface transmission coefficient $T_{\text{int}1}$ using different energy dispersion relations and different electron masses in metal and semiconductor parts [14]. However, we make further simplifications and assume that $T_{\text{int}1}$ is constant within the energy range where tunneling through the barrier is most efficient, and $T_{\text{int}2}$ is close to one according to the WKB approximation. In this case, interface transmission coefficient, $T_{\text{int}1}$, can be moved out of the integral in Eq. (1). Using the parabolic energy dispersion relation in the distribution function, $f_m(E)$, and neglecting $f_s(E)$, one can rewrite Eq. (1) as

$$j = \frac{A_\sigma^* T}{k_B} \int_{-E_F}^{\infty} T(E_x) \ln\left(1 + e^{-\frac{E_x}{k_B T}}\right) dE_x, \qquad (3)$$

where $A_\sigma^*$ is the modified Richardson constant, $k_B$ is the Boltzmann constant, $T$ is the lattice temperature, $E_x$ is the $x$-component of electron energy with respect to the Fermi level $E_F$, and $T(E_x)$ is the tunneling coefficient. Eq. (3) is similar to the standard formula utilized in electronics [12]. Though, it accounts for the Fermi-Dirac electron energy distribution in the metal contact. The conventional Richardson constant [12] should be corrected to include spin non-degeneracy and the interface transmission coefficient $T_{\text{int}1}$. Using Eq. (3), tunneling injection through the Schottky barrier can be efficiently modeled using the Monte Carlo method [22,15,23]. We have expanded the method described in [15] to account for the important aspects of 3D injection. For a given potential profile the energy distribution of electrons tunneled into the semiconductor during the time interval $\Delta t$ is calculated using eq. (3). We assume that the tunneling barrier is one dimensional, and therefore, it affects only the $x$-component of the electron energy. The initial states of the successfully tunneled electrons are generated based on the Fermi-Dirac energy distribution on the metal side of the barrier. In both, metal and semiconductor, energy dispersion relations are assumed to be parabolic. The spin state of the electrons injected is conserved during the tunneling and start to evolve under the influence of a spin-dependent Hamiltonian during the drift motion in the semiconductor channel of the device.

    To describe spin dynamics of injected electrons, we take into account the spin-orbit coupling due to the inversion asymmetry of GaAs crystal [16]. This interaction together with electron momentum scattering events produces rotational spin dephasing in

a system of spin-polarized electrons [30]. In the model, it is implemented as the spin rotation of each electron in the driving field during the electron free propagation between two scattering events. The initial states of electrons are obtained from the injection procedure. The profile of the Schottky barrier is determined by the self-consistent solution of charge carrier motion and the Poisson equation. To describe spin of a single electron, we use a single electron spin density matrix [31,32] representation that is equivalent to the spin polarization vector description [33,34]. Details of the simulation procedure applied for spin transport in low dimensional structures can be found in [31]. For the electron wave vector in the vicinity of points of high symmetry within the Brillouin zone, an analytical form for corrections to the effective mass Hamiltonian due to the spin-orbit interaction has been derived in [16]. Near the bottom of the valleys, the spin-orbit Hamiltonians can be written in the following form [35]

$$H_\Gamma = \alpha_\Gamma \left( k_x \sigma_x (k_y^2 - k_z^2) + k_y \sigma_y (k_z^2 - k_x^2) + k_z \sigma_z (k_x^2 - k_y^2) \right) \quad (4)$$

in the Γ-valley,

$$H_L = \alpha_L / \sqrt{3} \left( \sigma_x (k_y - k_z) + \sigma_y (k_z - k_x) + \sigma_z (k_x - k_y) \right) \quad (5)$$

in the L-valley located along the [111] direction in the crystallographic axes, and

$$H_X = \alpha_X \left( \sigma_z k_z - \sigma_y k_y \right) \quad (6)$$

in the X-valley located along [100] direction. In Eqs. (4)-(6), $\sigma_\alpha$ are the spin Pauli matrixes [29], $k_\alpha$ the components of an electron wave vector, and $\alpha_\Gamma$, $\alpha_L$ and $\alpha_X$ the spin-orbital coupling coefficients. Explicit forms of spin-orbit Hamiltonians for other equivalent valleys of L and X symmetries can be obtained by appropriate rotation of the coordinate system. In the model, we assume that one of the appropriate equations in Eqs. (4)-(6) can be applied to electrons in a given valley. However, it should be noticed that these equations are correct only near the bottoms of valleys. The spin-orbit coupling coefficients utilized in simulations are $\alpha_\Gamma = 28$ eVÅ$^3$ [36], $\alpha_L = 0.27$ eVÅ [37], and $\alpha_X = 0.087$ eVÅ [36]. We would like to emphasize that due to lack of information on spin coupling parameters in upper valleys their actual values can be different.

**Simulation Results and Discussions**

We simulated steady-state injection of spin-polarized electrons into the device shown in Fig. 1 for $T = 4.2 - 350$ K, and $V_{\text{bias}} = 1 - 3$ V. The concentration and average energy profiles of electrons injected at $T = 4.2$ K and $V_{\text{bias}} = 2$ V are shown in Fig. 2 (a)-(b). Near the metal/semiconductor interface, electrons injected into semiconductor are strongly accelerated by the interface electric field. On a distance about 20 – 30 nm from the interface, the intervalley scattering mechanism becomes dominant, and electrons are redistributed among the valleys. As it has been demonstrated in previous studies [20], the electron population becomes even higher in upper valleys than in the Γ-valley. Scattering of electrons back to the Γ-valley becomes evident after traveling for a distance on the order of several hundred nanometers. These profiles remain similar for the whole ranges of voltage and temperature, though actual values of concentration and energy change. At low temperature, practically, all the electrons are injected by tunneling near the Fermi level in the metal contact, while at temperatures ~ 50 – 70 K thermal assisted tunneling starts to play a role. The contribution of the thermionic emission becomes noticeable only at temperatures above 150 K.

To characterize spin properties of injected current, we use electron spin polarization, defined as

$$S_\alpha = \sum_i \text{Tr}(\sigma_\alpha \rho_\sigma^i) / \sum_i \text{Tr}(\rho_\sigma^i), \qquad (7)$$

and current spin polarization

$$P_\alpha = \sum_i v_x^i \text{Tr}(\sigma_\alpha \rho_\sigma^i) / \sum_i v_x^i \text{Tr}(\rho_\sigma^i), \qquad (8)$$

where $\rho_\sigma^i$ is the single electron spin density matrix [29] and $v_x^i$ is the $x$-component of velocity of the $i$-th particle. In Eqs. (7)-(8), sums are taken over all the injected particles located within a small volume $d^3x$ near the position $\mathbf{x}$. Eqs. (7)-(8) are similar to standard definitions of particle and flux polarization utilized in literature [1,38]. In general, electron spin polarization is easier to measure experimentally, for example, using the oblique Hanle effect [11]. However, in a device, electrons injected represent only a small fraction of the total number of carriers. In the structure studied, Fig. 1, total electron spin polarization is negligible except for the depletion region. Current spin polarization, Eq. (8), by definition, cancels the effect of background electrons in the device channel that do not contribute to the current. However, it is not clear how this characteristic can be measured. One of the possible methods is discussed in [39].

In simulations, we observe that both characteristics of spin polarization decay very rapidly in a length scale of the order of 50 – 100 nm. Simulated profiles at $T = 4.2$ K and $V_{\text{bias}} = 2$ V are shown in Fig. 3. Spin dynamics, controlled by spin-orbit coupling, Eqs. (4)-(6), and electron momentum distributions are different in different valleys. With the given set of transport parameters, we obtain that spin characteristics in upper valleys decay about twice faster rather than those in the Γ-valley for the temperature and applied bias ranges, $T = 4.2 – 350$ K and $V_{\text{bias}} = 1 – 3$ V. In the Γ-valley, both electron and current spin polarization profiles can be fitted by $\exp(-(x/x_0)^\alpha)$, where coefficient $\alpha \sim 2$-3, rather than by the linear exponential decay function with $\alpha = 1$. This dependence cannot be explained within the drift-diffusion approximation [38,40]. To characterize spin dephasing, we use electron and current spin scattering lengths which are defined as the lengths where electron spin and current spin polarizations decay to 1/*e* of their initial values, respectively. With the temperature increment, both spin scattering lengths decrease monotonically, Fig 4. At room temperature spin scattering lengths in the Γ-valley are more than twice shorter than at *T* = 4.2 K. Evident bias voltage dependence of spin scattering lengths is observed for electrons in the Γ-valley only, Fig. 5. Similarly to the studies using drift-diffusion models [38,41,40], our results show that at higher bias electron spin polarization penetrates more deeply into the semiconductor. However, this dependence is weak. For the current spin scattering length, we observe a reverse tendency. It becomes shorter at higher bias.

Signatures of hot spin-polarized carrier transport in structures with Schottky barriers have been observed recently in [42]. Our simulations confirm that spin-polarized transport can be detected in spin-LED structures. In more detailed comparison of simulated results with experimental data on spin injection in spin-LEDs we find that the model provides qualitatively consistent spin scattering length scales with some overestimation of the spin dephasing effect. The latter can be due to different setups in the simulations and experimental study, and also due to not well matched transport parameters. Transport parameters within the range accessible in literatures gives variation in electron and current spin scattering lengths of the order of 10 – 15 %. The most crucial are spin-orbit coupling coefficients. For the L and X valleys, only theoretical estimations [36,37] are available. For the Γ-valley, the measured value of spin-orbit coupling coefficient corresponds to its minimum. However, in the simulations, electrons in this valley possess high kinetic energy, Fig.2(b). In this case, the coupling coefficient should be modified. We have also accounted for spin rotation in a homogeneous external magnetic field that is utilized in experiments with spin-LEDs. For simplicity we used isotropic g-factors $g_\Gamma^* = -0.44$ [43], $g_L^* = 2$ [44], and $g_X^* = 2$. However, we found that the effect is negligible for the magnetic fields up to *H* = 2 T.

**Conclusion**

In conclusion, we simulated injection of spin-polarized electrons through a Schottky barrier into a GaAs device channel. For the whole temperature range, T = 4.2 – 350 K, and the applied bias range, $V_{bias}$ = 1 – 3 V, we observed hot-electron transport features. Electrons injected through the barrier quickly redistributed among the semiconductor valleys. On the length scale of the order of 200 – 300 nm from the metal/semiconductor interface, upper valleys (X and L) are more populated by the injected electrons than the Γ-valley. Electron and current spin polarizations injected decay in a very short length scale of the order of 50-100 nm. Within the model developed, spin polarization characteristics decay faster in upper valleys.

*Acknowledgements.* We are thankful to Robert Mallory, Athos Petrou, Vladimir Privman, and Mesut Yasar for useful discussions. The research was supported by the National Science Foundation, grant DMR-0121146. The work of S.S. is also supported by DMR-0403465

**References:**

1. I. Zutic, J. Fabian, S. Das. Sarma, Spintronics: Fundamentals and applications, Rev. Mod. Phys. **76**, 323 (2004)

2. B. T. Jonker, Progress toward electrical injection of spin-polarized electrons into semiconductors, Proc. IEEE **91**, 727 (2003)

3. K. H. Ploog, Spin injection in ferromagnet-semiconductor heterostructures at room temperature, J. Appl. Phys. **91**, 7256 (2002)

4. R. Fiederling, M. Keim, G. Reuscher, W. Ossau, G. Schmidt, A. Waag, L. W. Molenkamp, Nature **402**, 787 (1999)

5. J. C. Egues, Phys. Rev. Lett. **80**, 4578 (1998)

6. G. Schmidt, D. Ferrand, L. W. Molenkamp, A. T. Filip, B. J. van Wees, Fundamental obstacle for electrical spin injection from a ferromagnetic metal into a diffusive semiconductor, Phys. Rev. B **62**, R4790 (2000)

7. E. I. Rashba, Theory of electrical spin injection: Tunnel contacts as a solution of the conductivity mismatch problem, Phys. Rev. B **62**, R16267 (2000)

8. J. D. Albrecht, D. L. Smith, Spin-polarized electron transport at ferromagnet/semiconductor Schottky contacts, Phys. Rev. B **68**, 035340 (2003).

9. S. F. Alvarado, P. Renaud, Observation of spin-polarized-electron tunneling from a ferromagnet into GaAs, Phys. Rev. Lett. **68,** 1387 (1992)

10. A. T. Hanbicki, B. T. Jonker, G. Itskos, G. Kioseoglou, A. Petrou, Efficient electrical spin injection from a magnetic metal/ tunneling barrier contact into a semiconductor, Appl. Phys. Lett. **80**, 1240 (2002)

11. V. F. Motsnyi, P. Van Dorpe, W. Van Roy, E. Goovaerts, V. I. Safarov, G. Borghs, J. De Boeck, Optical investigation of electrical spin injection into semiconductors, Phys. Rev. B **68**, 245319 (2003)

12. S. M. Sze, Physics of semiconductor devices, (John Wiley & Sons, New York 1981)

13. C. Adelmann, X. Lou, J. Strand, C. J. Palmstrom, P. A. Crowell, Spin injection and relaxation in ferromagnet-semiconductor heterostructures, Phys. Rev. B **71**, 121301(R) (2005)

14. V. V. Osipov, A. M. Bratkovsky, Efficient nonlinear room-temperature spin injection from ferromagnets into semiconductors through a modified Schottky barrier, Phys. Rev. B **70**, 205312 (2004)

15. M. Shen, S. Saikin, M.-C. Cheng, Monte Carlo modeling of spin injection through a Schottky barrier and spin transport in a semiconductor quantum well, J. Appl. Phys. **96**, 4319 (2004)

16. G. Dresselhaus, Spin-orbit coupling effects in Zinc Blende structures, *Phys. Rev.*, **100**, 580 (1955)

17. A.T. Hanbicki, O.M.J. van t Erve, R. Magno, G. Kioseoglou, C.H. Li, B.T. Jonker, G. Itskos, R. Mallory, M. Yasar, A. Petrou, Analysis of the Transport Process Providing Spin Injection through an Fe/AlGaAs Schottky Barrier, Appl. Phys. Lett. **82**, 4092 (2003)

18. J. R. Waldrop, Schottky-barrier height of ideal metal contact to GaAs, Appl. Phys. Lett. **44**, 1002 (1984)

19. C. Moglestue, Monte Carlo simulation of semiconductor devices, (Chapman & Hall, New York 1993)

20. J. Pozela, A. Reklaitis, Solid State Electronics **23**, 927 (1980)

21. I. Appelbaum, V. Narayanamurti, Monte Carlo calculations for metal-semiconductor hot-electron injection via tunnel-junction emission, Phys. Rev. B **71**, 045320 (2005)

22. L. Sun, X.Y. Liu, M. Liu, G. Du, R.Q. Han, Semiconductor Science and Technology, **18**, 576 (2003)

23. Y. Y. Wang, M. W. Wu, Schottky-barrier induced spin dephasing in spin injection, Phys. Rev. B **72**, 153301 (2005)

24. For scattering formulas see M. V. Fischetti, S. E. Laux, DAMOCLES Theoretical manual (IBM Research Division, Yorktown heights 1995)

25. X.–G. Zhang, W. H. Butler, Band structure, evanescent states, and transport in spin tunnel junctions, J. Phys.: Condens. Matter **15**, R1603 (2003)

26. J. Callaway, C. S. Wang, Energy bands in ferromagnetic iron, Phys. Rev. B **16**, 2095 (1977)

27. W. Kohn, Effective mass theory in solids from a many-particle standpoint, Phys. Rev. **105**, 509 (1957)

28. C. B. Duke, Theory of metal-barrier-metal tunneling, *in Tunneling phenomena in solids* (Plenum Press, New York, 1969)

29. E. Merzbacher, Quantum Mechanics, (John Wiley & Sons, New York 2004)

30. M. I. D.yakonov, V. I. Perel, Soviet Phys. JETP **33**, 1053 (1971)

31. S. Saikin, M. Shen, M.-C. Cheng, V. Privman, J. Appl. Phys. **94,** 1769 (2003)

32. S. Pramanik, S. Bandyopadhyay, and M. Cahay, Spin dephasing in quantum wires, Phys. Rev. B **68**, 075313 (2003)

33. A. A. Kiselev and K. W. Kim, Progressive suppression of spin relaxation in two-dimensional channels of finite width, Phys. Rev. B **61**, 13115 (2000)

34. Y. V. Pershin and V. Privman, Slow spin relaxation in two-dimensional electron systems with antidotes, Phys. Rev. B **69**, 073310 (2004)

35. E. L. Ivchenko, G. E. Pikus, Superlattices and other Heterostructures. Symmetry and Optical Phenomena. In Solid-State Sciences vol. 110 (Springer-Verlag, Berlin 1997)

36. M. Cardona, N. E. Christensen, G. Fasol, Relativistic band structure and spin-orbit splitting of zinc-blende-type semiconductors, Phys. Rev. B **38**, 1806 (1988)

37. J.-M. Jancu, R. Scholz, G. C. La Rocca, E. A. de Andrada e Silva, P. Voisin, Giant spin splittings in GaSb/AlSb L-valley quantum well, Phys. Rev. B **70**, 121306(R) (2004)

38. Z. G. Yu, M. E. Flatte, Spin diffusion and injection in semiconductor structures: Electric field effects, Phys. Rev. B **66**, 235302 (2002)

39. A. Di Lorenzo, Y. V. Nazarov, Full Counting Statistics of Spin Currents, Phys. Rev. Lett. **93**, 046601 (2004)

40. S. Saikin, Drift-diffusion model for spin-polarized transport in a non-degenerate 2DEG controlled by a spin-orbit interaction, J. Phys.: Condens. Matter **16**, 5071 (2004)

41. I. Zutic, J. Fabian, S. Das Sarma, Spin-polarized transport in inhomogeneous magnetic semiconductors: theory of magnetic/nonmagnetic p-n junctions, Phys. Rev. Lett. **88**, 066603 (2002)

42. R. Mallory, M. Yasar, G. Itskos, A. Petrou, G. Kioseoglou, A.T. Hanbicki, C.H. Li, O.M.J. van 't Erve, B.T. Jonker, M. Shen, S. Saikin, preprint (2005)

43. O. Madelung, Semiconductors – Basic data, (Springer-Verlag, Berlin 1996)

44. F. A. Baron, A. A. Kiselev, H. D. Robinson, K. W. Kim, K. L. Wang, E. Yablonovitch, Manipulating the L-valley electron g factor in Si-Ge heterostructures, Phys. Rev. B **68**, 195306 (2003)

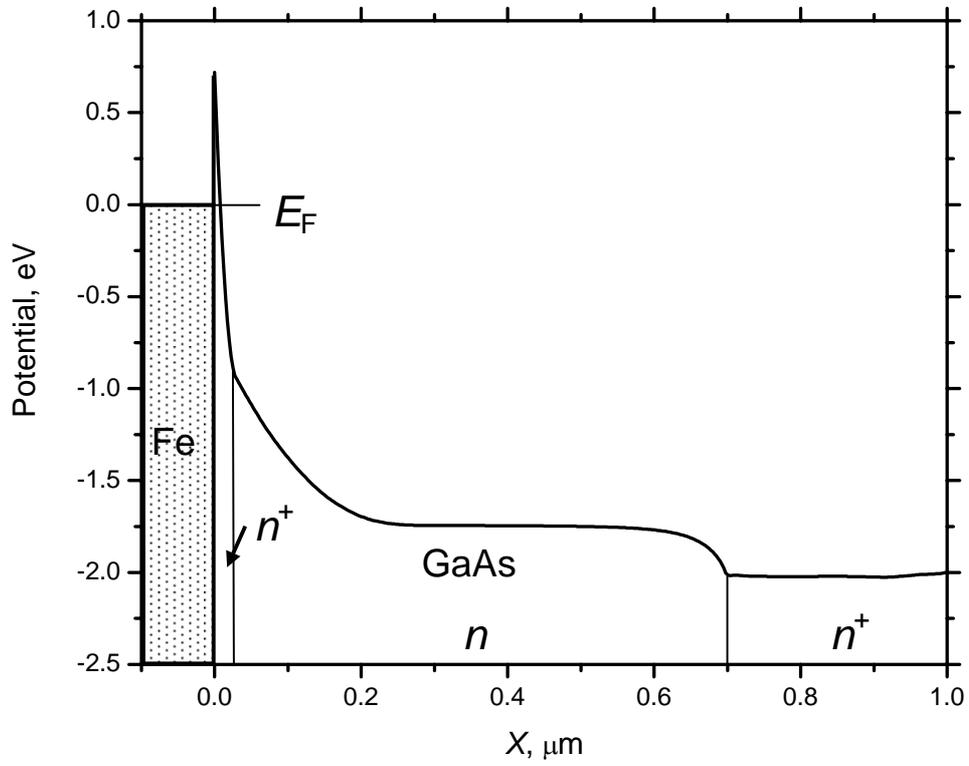

**Figure 1.** The potential profile together with the doping profile of the simulated structure at $T = 4.2$ K and $V_{bias} = 2$ V.

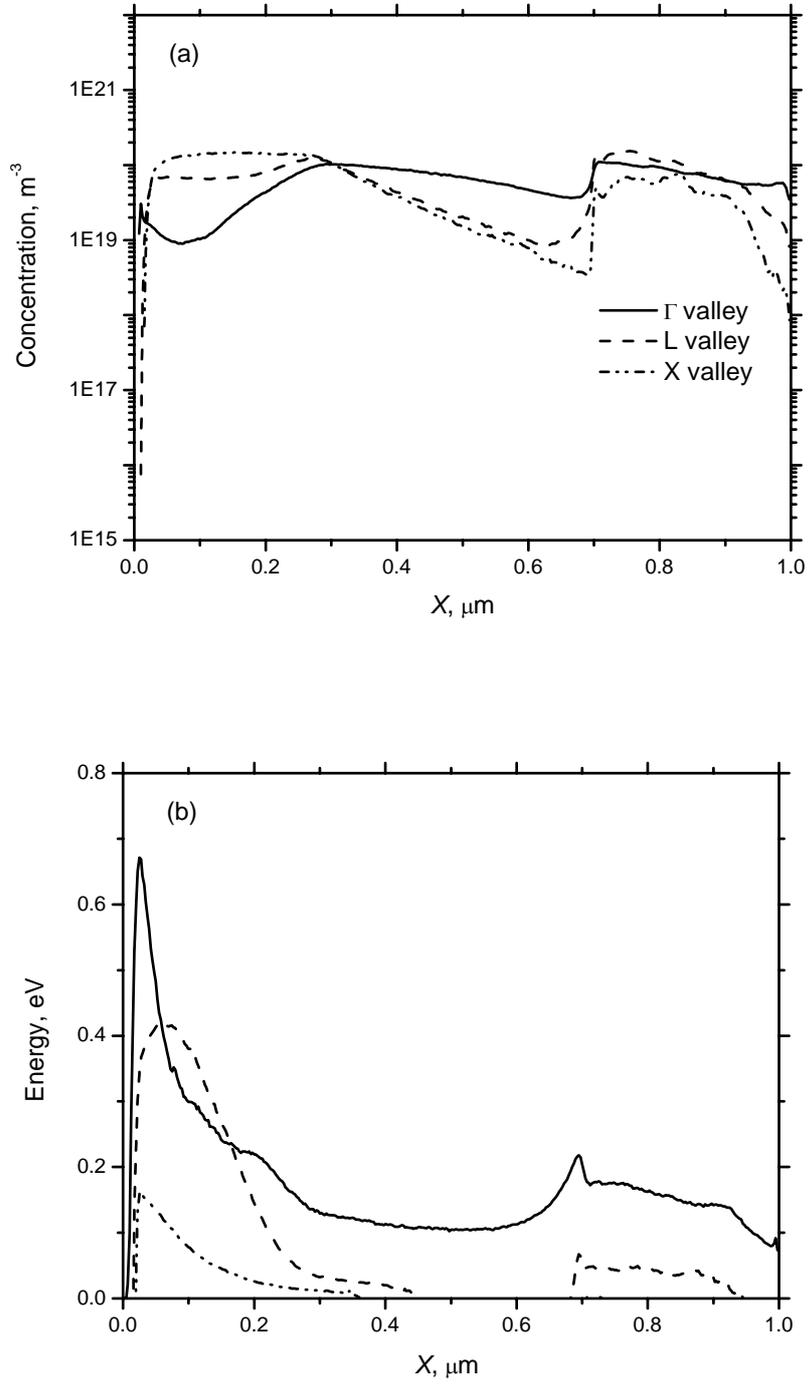

**Figure 2**. Populations of electrons in different valleys injected into the device (a), average energies of injected electrons (b) at $T$ = 4.2 K, and $V_{\text{bias}}$ = 2 V.

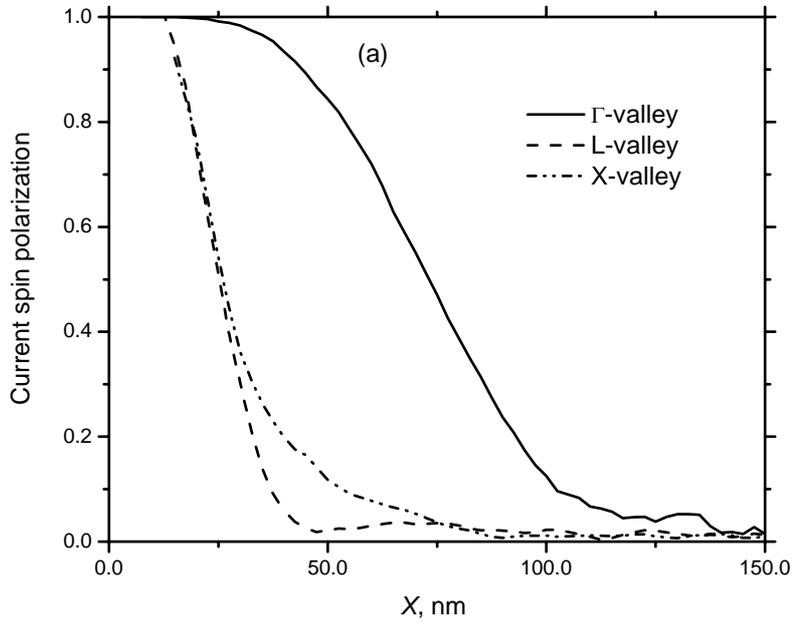

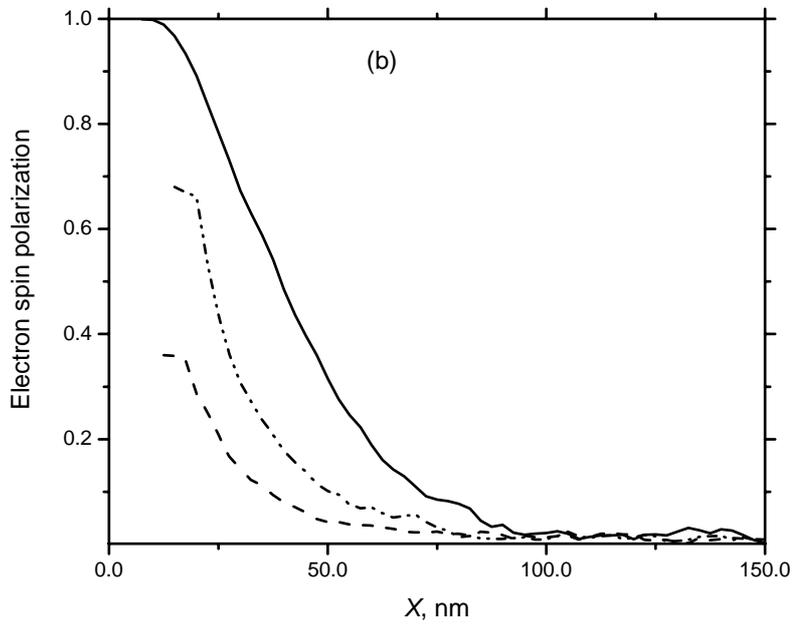

**Figure 3**. Current spin polarization profile (a), and electron spin polarization profile (b) simulated at $T = 4.2$ K and $V_{bias} = 2$ V.

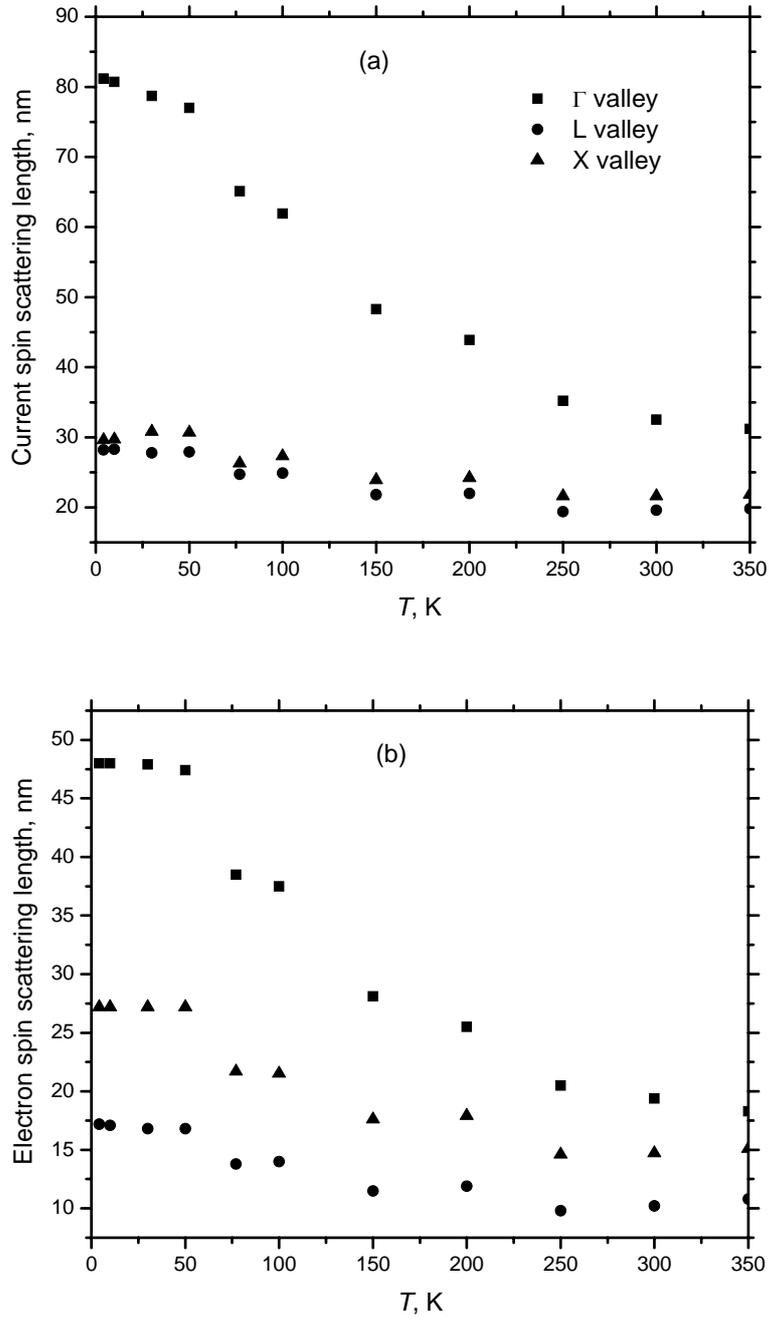

**Figure 4**. Temperature dependences of current spin scattering length (a), and electron spin scattering length (b) at $V_{bias} = 2$ V.

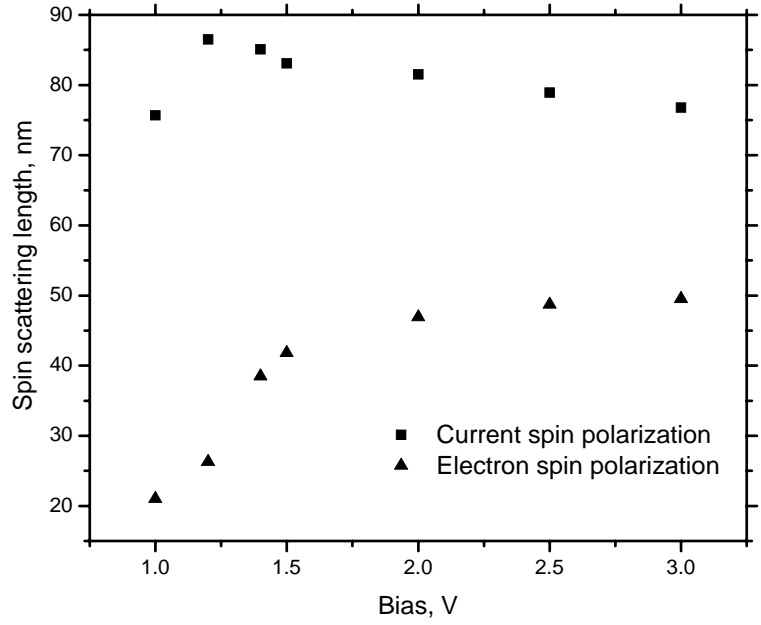

**Figure 5**. Current spin and electron spin scattering lengths in the Γ-valley as functions of applied bias at $T = 4.2$ K.